\documentclass[letterpaper,10pt]{article} 
\usepackage{indentfirst} 
\usepackage{opticameet3} 
\usepackage{upgreek}
\usepackage{graphicx}
\graphicspath{ {./} }
\usepackage[justification=centering]{caption}

\newcommand\authormark[1]{\textsuperscript{#1}}

\usepackage{amsmath,amssymb}
\usepackage[colorlinks=true,bookmarks=false,citecolor=blue,urlcolor=blue]{hyperref} 

\begin{document}

\title{The effect of substrate temperature on cadmium telluride films in high temperature vapor deposition process}


\author{Wenxiong Zhao\authormark{*} and Yufeng Zhang\authormark{*}}

\address{The key laboratory of solar thermal and photovoltaic system, institute of electrical engineering, CAS, Beijing 10019, China.\\}

\email{\authormark{*}These authors contributed equally to this work. wxzhao@mail.iee.ac.cn } 

\begin{abstract}

Physical vapor high-temperature deposition of CdTe thin films is one of the main methods for preparing high-efficiency CdTe solar cells, but high-temperature deposition also has an impact on the internal structure of the film. The difference in thermal expansion coefficients between the substrate and CdTe leads to the generation of internal stress in the CdTe thin film during the cooling process. In this work, we prepared thin films with different substrate temperatures using a homemade GVD device, and observed by SEM that the crystallization quality of the film gradually improved with the increase of substrate temperature, but accompanied by the shift of XRD peak position. We calculated the internal stress situation of the film by the shift amount, and the possible causes of stress generation were speculated by the results of TEM and SAED to be the combined effects of the different thermal expansion coefficients between the substrate and the film and the stacking fault defects inside the film. 
\end{abstract}

\section{Introduction}

Cadmium telluride is an important photovoltaic material, with advantages such as ideal band gap width, high light absorption rate, high conversion efficiency and low cost\cite{ref1}. The quality of cadmium telluride thin films largely determines the quality of cadmium telluride solar cells. How to prepare high-quality cadmium telluride thin films is one of the key issues that relevant scholar’s study. In the current high-efficiency cadmium telluride thin film solar cells, close-spaced sublimation and vapor transport deposition have been proven to be methods for preparing high-quality cadmium telluride thin films\cite{ref2,ref3}, and some production lines of cadmium telluride thin film solar cells have already used these methods for production. These film deposition methods are all high-temperature deposition methods, that is, CdTe is deposited on the substrate surface when the substrate is heated at a high temperature. This high-temperature deposition method can significantly increase the grain size of the film, improve the density of the film, reduce the probability of carrier recombination at the grain boundary, help the carrier to be effectively separated and collected, and ultimately enable the solar cell efficiency to be effectively improved\cite{ref4}.

But high-temperature deposition process inevitably has some problems. The high-temperature state will intensify the thermal motion of atoms, making the atoms in the crystal more likely to undergo displacement, diffusion, and rearrangement, these problems will cause more defects in the film. And the thermal expansion coefficients of the substrate and CdTe film are different, the difference in shrinkage during the cooling process will cause internal stress in the film. Therefore, it is very necessary to explore the situation of CdTe films deposited at high temperatures.

In this work, a self-made guided vapor deposition(GVD) device was used to prepare CdTe thin films. Under the condition of not using carrier gas, CdTe vapor was self-diffused and sprayed downward in the heated sublimation tube, and finally CdTe thin films were deposited on the surface of different temperature substrates facing up. The effect of high-temperature vapor deposition process on the morphology, crystal plane orientation, internal stress, and defects of CdTe thin films was explored.

\section{Experimental method}

\begin{figure}[htp]
    \centering
    \includegraphics[width=11cm]{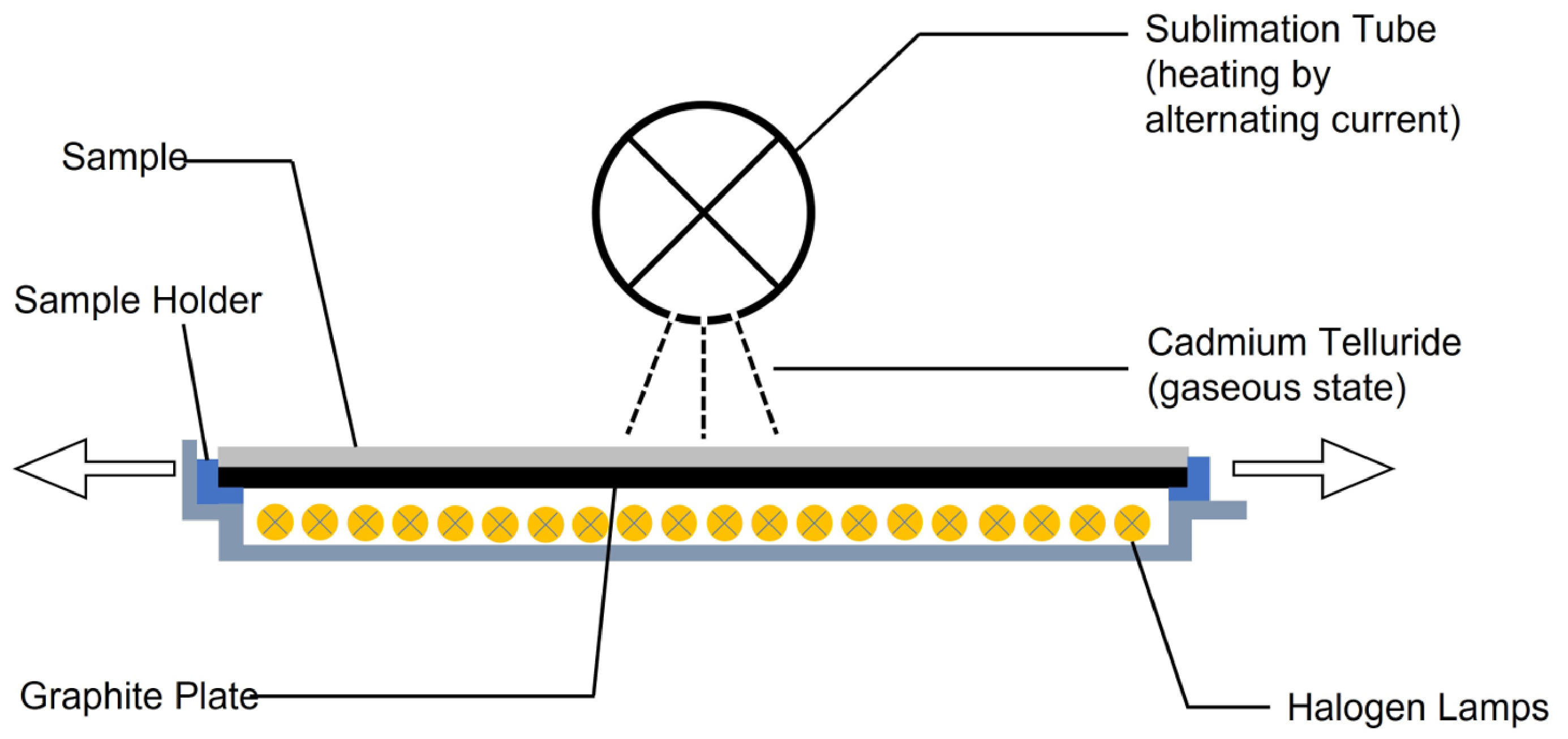}
    \caption{Schematic diagram of self-made GVD device.}
    \label{fig.1}
\end{figure}

High-purity cadmium telluride particles (99.99\%) were used as the deposition material, and cadmium telluride thin films were deposited on glass coated with Sn$\rm O_2$:F by a self-made vapor-guided deposition method (Fig.\ref{fig.1}). The temperature of the sublimation source was heated to 830°C to make the cadmium telluride vapor spray out of the sublimation guide pipe quickly. Below was a heated tray carrying the substrate, and the deposition rate of cadmium telluride thin films was controlled by controlling the tray temperature and moving speed. Before deposition, the chamber was pumped to 3×$\rm 10^{-3}$Pa and nitrogen was introduced to maintain the pressure of 1Pa during deposition. The substrate temperatures for cadmium telluride deposition were set at 490°C, 520°C, 550°C, and 580°C, respectively. These temperatures were calibrated by contacting the substrate with a K-type thermocouple before deposition. The thickness of the prepared films was controlled at about 2.7$\upmu$m, and the crystal structure of the films was characterized by X-ray diffraction (XRD), all XRD test results were calibrated by the peak position offset of single crystal silicon fragments samples. The morphology of the films was characterized by scanning electronic microscopy (SEM). The sample was cut by focused ion beam (FIB), and the twin and dislocation situations were characterized by transmission electron microscopy (TEM) and selected area electron diffraction (SAED).

\section{Results and discussions}
\begin{figure}[htp]
    \centering
    \includegraphics[width=13cm]{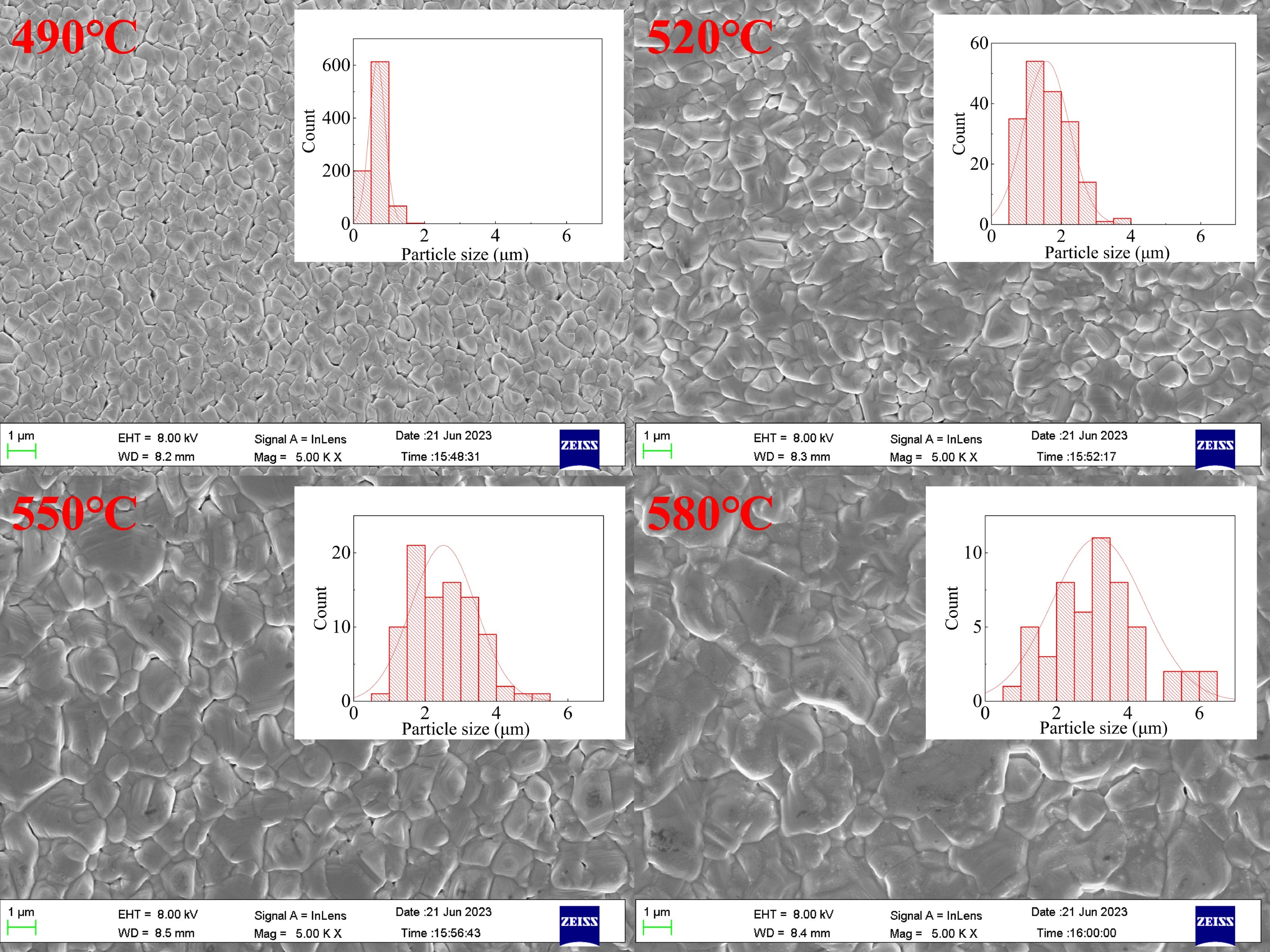}
    \caption{SEM surface morphology and grain size distribution of CdTe thin films deposited at different substrate temperatures.}
    \label{fig.2}
\end{figure}
The surface morphology of CdTe thin films prepared at different substrate temperatures observed by SEM is shown in Fig.\ref{fig.2}. With the gradual increase of substrate temperature, it can be clearly observed that the deposited films gradually become denser, and the grain boundaries gradually become less obvious. The grain size was statistically analyzed by Nano Measurer 1.2 software, and the grain size increased gradually with the increase of substrate temperature.

From the perspective of physical chemistry\cite{ref5}, the difference of phase transition free energy per unit volume of solid phase in the condensation process is shown as equation (1):
\begin{eqnarray}
{\Delta G_V} = -\frac{{kT}}{\Omega} ln(1+S)
\end{eqnarray}
where $\Omega$ is the atomic volume, $S$ is the supersaturation of gas phase, the film deposition process means that there is supersaturation phenomenon in gas phase, that is, $\Delta G_V$ \textless 0, assuming that the radius of new phase nucleus is $r$, then the bulk free energy will change $(4/3)\pi r^3\Delta G_V$ at nucleation, the interfacial energy between solid and gas will increase $4\pi r^2 \gamma$, where $\gamma$ is the interfacial energy per unit area, based on the above obtain the overall free energy change $\Delta G$ of deposition process as described by equation (2):
\begin{eqnarray}
{\Delta G} = \frac{4}{3}{\pi r^3 \Delta G_V + 4\pi r^2 \gamma }
\end{eqnarray}

In the process of gradually increasing the substrate temperature, to prepare CdTe thin films with the same thickness, this work had to gradually extend the time spent in deposition process. As shown in equation (3):
\begin{eqnarray}
{\tau _a} = \tau_0 exp(\frac{E_d}{RT} )
\end{eqnarray}
where $\tau_a$ represents the average adsorption time of molecules on substrate surface; $\tau_0$ is the inverse of surface particle vibration frequency; $E_d$ is the desorption activation energy of material; $R$ is gas constant; $T$ is thermodynamic temperature. When substrate temperature $T$ gradually increases, obviously surface particle vibration frequency gradually increases, that is, $\tau_0$ decreases, therefore average adsorption time $\tau_a$ of molecules on substrate surface decreases, CdTe molecules are more difficult to stay on substrate surface, reflecting to actual situation is that deposition rate of thin film gradually decreases. And phase transition free energy of gas phase condensing into solid phase can be expressed in form of equation (4):
\begin{eqnarray}
{\Delta G_V} = -\frac{kT}{\Omega }ln\frac{R_a}{R_e} 
\end{eqnarray}
where $R_e$ is equilibrium evaporation rate of solid phase nucleus at temperature $T$, and $R_a$ is actual deposition rate. When thin film deposits, $\Delta G_V$ \textless 0, differentiate equation (4) with respect to $r$, obtain condition for free energy $\Delta G=0$ as $r^*=-(2\gamma/\Delta G_V)$, $r^*$can be considered as minimum solid phase nucleus radius that can balance existence, when deposition substrate temperature $T$ rises at same time, accompanied by decrease of deposition rate $R_a$, $\Delta G_V$  gradually decreases, minimum solid phase nucleus radius that can balance existence will gradually increase, this reflects in this work as process of CdTe thin film grain enlargement.

\begin{figure}[htp]
    \centering
    \includegraphics[width=13cm]{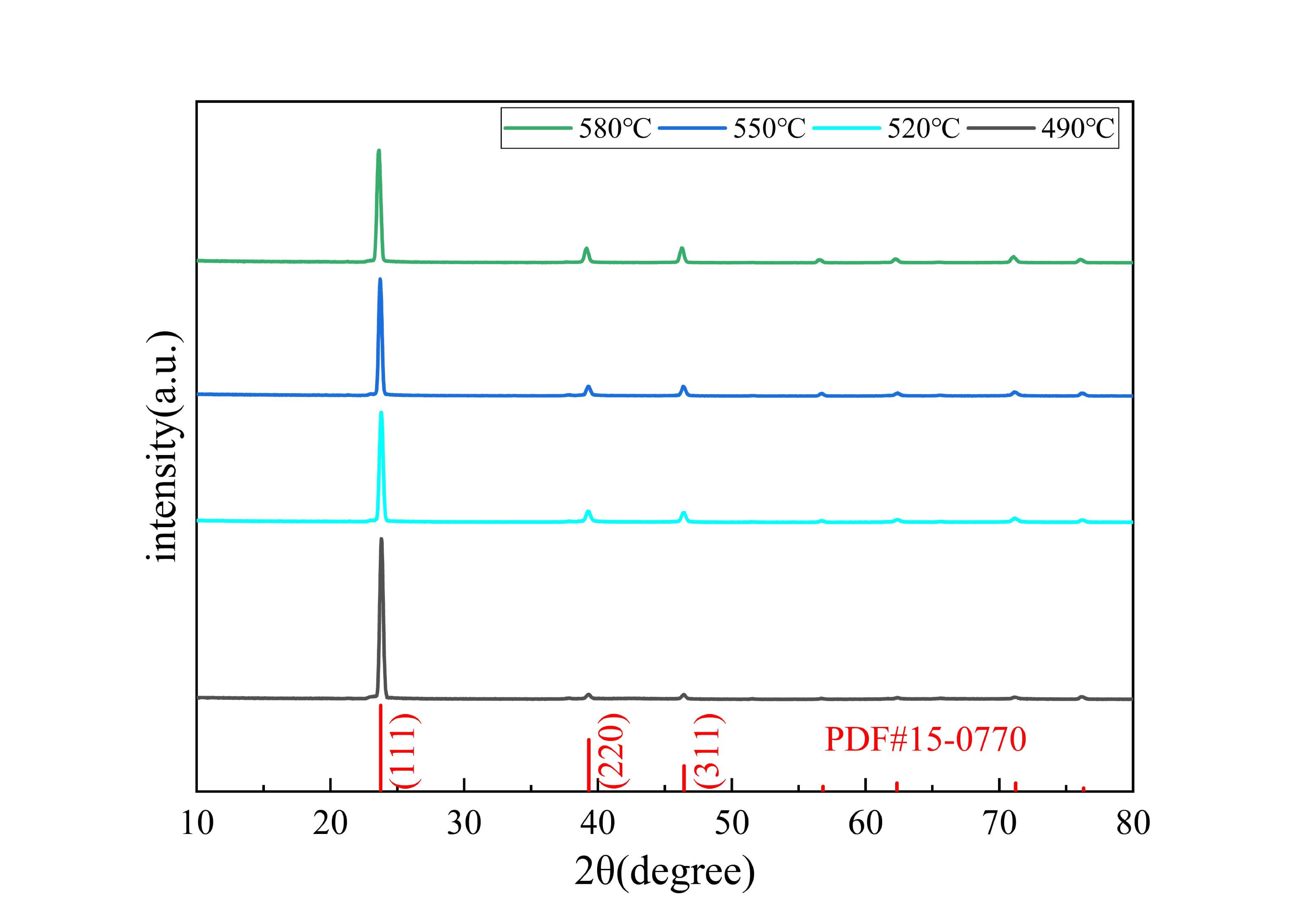}
    \caption{XRD diagram of CdTe thin films grown at different substrate temperatures.}
    \label{fig.3}
\end{figure}

The XRD patterns of CdTe thin films grown at different substrate temperatures are shown in Fig. \ref{fig.3}. When the substrate temperature during deposition was 490°C, the test results showed that the grains were almost all preferentially grown along the (111) plane, and other diffraction peaks were almost invisible. With the further increase of the substrate temperature during deposition, the diffraction peak of the (111) plane was reduced, while the diffraction peaks of the (220) and (311) planes gradually increased, and the film gradually tended to random grain orientation. The possible reason is that the nucleation barrier of cubic CdTe in the (111) direction is lower than that in other directions, which makes it possible to quickly form many grains at low temperature. However, the nucleation barrier in other directions is higher, resulting in low grain density and few grains formed at low temperature. With the increase of substrate temperature, the nuclei in other directions gradually grow into grains, and the corresponding diffraction peak intensity also gradually increases.

The texture coefficient $TC(h_i k_i l_i)$ can reflect the orientation of the film, which can be calculated by equation (5)\cite{ref6}:
\begin{eqnarray}
{TC(h_ik_il_i)} = \frac{\frac{I(h_ik_il_i)}{I(h_0k_0l_0)}}{\frac{1}{N}[ {\textstyle \sum_{1}^{N}}\frac{I(h_ik_il_i)}{I(h_0k_0l_0)} ] }
\end{eqnarray}
where $I(h_i k_i l_i)$ is the diffraction peak intensity of the (hkl) plane measured by XRD, $I(h_o k_o l_o)$ is the standard intensity of the specific (hkl) plane on the PDF card (this work uses the cubic CdTe PDF\#15-0770), and $N$ is the number of diffraction peaks. The texture coefficients of the samples under different conditions were calculated based on the XRD results of each crystal plane, as shown in Table 1:
\begin{table}[htb]
 \centering \caption{Texture coefficients of each plane of CdTe thin films deposited at different temperatures.}
\begin{tabular}{ccccc}
    \hline
     & 490℃ & 520℃ & 550℃ & 580℃\\
    \hline
    (111) & 3.48 & 2.35 & 2.25 & 1.79 \\
    (220) & 0.19 & 0.30 & 0.32 & 0.39 \\
    (311) & 0.36 & 0.65 & 0.64 & 0.78 \\
    (400) & 0.47 & 0.69 & 0.89 & 0.88 \\
    (331) & 0.41 & 0.66 & 0.64 & 0.65 \\
    (422) & 0.55 & 0.92 & 0.82 & 1.03 \\
    \hline
   \end{tabular}
    \end{table}
from the results of the texture coefficients, with the increase of the substrate temperature during deposition, the film gradually changes from preferential orientation of the (111) plane to random orientation. This kind of random orientation film usually appears after high-temperature deposition or Cd$\rm Cl_2$ treatment. The random orientation absorber layer can increase the scattering and capture of light in it, which is helpful for light absorption\cite{ref7,ref8}.

\begin{figure}[htp]
    \centering
    \includegraphics[width=13cm]{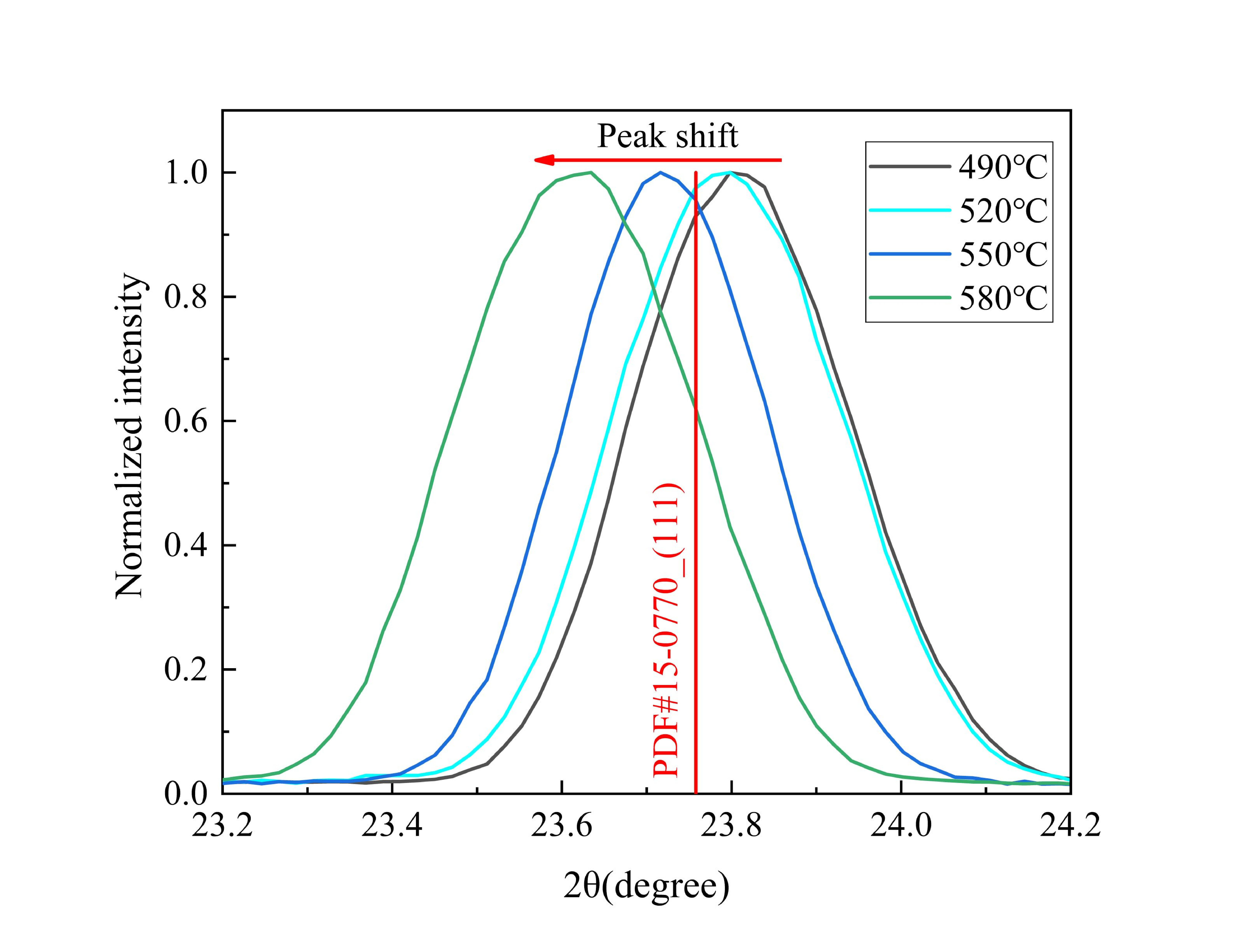}
    \caption{Normalized results of the (111) peak position of the XRD diagram of CdTe thin films grown at different substrate temperatures.}
    \label{fig.4}
\end{figure}

After normalizing the (111) plane in Fig.\ref{fig.3}, the peak positions were compared, as shown in Fig.\ref{fig.4}. The peak positions gradually shift to small angles, indicating that the longitudinal interplanar spacing of the film gradually increases with the increase of substrate temperature. According to the offset of different peak positions in the XRD results, the change of interplanar spacing $\Delta d$ can be calculated by using Bragg’s equation, which is:
\begin{eqnarray}
{2dsin\theta } = {n\lambda} 
\end{eqnarray}
where $d$ is the interplanar spacing, $\theta$ is the incident angle, $n$ is the diffraction order, and $\lambda$ is the incident wavelength. If the position of the diffraction peak changes, the corresponding $\theta$ value will also change, resulting in the change of d value. Assuming that the standard diffraction peak position is $\theta_0$, the corresponding interplanar spacing is $d_0$, the current diffraction peak position is $\theta_1$, and the corresponding interplanar spacing is $d_1$, then:
\begin{eqnarray}
{d_0sin\theta _0 } = {d_1sin\theta_1} = {n\lambda }
\end{eqnarray}
therefore, the change of interplanar spacing can be calculated by the following formula:
\begin{eqnarray}
{\Delta d} = {d_1-d_0} = \frac{d_0(sin\theta _0-sin\theta _1)}{sin\theta _1} 
\end{eqnarray}
the strain values of different CdTe thin films after cooling can be calculated by the formula:
\begin{eqnarray}
{\varepsilon } = \frac{\Delta d}{d_0} 
\end{eqnarray}
the film stress can be calculated based on the biaxial strain model, where the formula applicable to the cubic phase is\cite{ref9}:
\begin{eqnarray}
{\sigma } = \frac{6C_{44}(C_{11}+2C_{12})}{C_{11}+2C_{12}+4C_{44}} \varepsilon \end{eqnarray}
for CdTe, $C_{11}$=53.5GPa, $C_{12}$=36.5 GPa, $C_{44}$=19.9 GPa, these elastic constants have been reported in other paper\cite{ref10}. According to equations (9) and (10), the transverse stress of the films deposited under different conditions was roughly estimated, and the specific results are shown in Table 2.
\begin{table}[htb]
 \centering \caption{Transverse stress estimation of CdTe films at different temperatures.}
\begin{tabular}{ccc}
    \hline
    Sample & (111) Crystal plane diffraction peak position (°) & Thin film transverse stress estimation (MPa)\\
    \hline
    490℃ & 23.818 & 65.19 \\
    520℃ & 23.798 & 43.46 \\
    550℃ & 23.716 & -45.65 \\
    580℃ & 23.634 & -134.82 \\
    \hline
   \end{tabular}
    \end{table}
according to the calculation results, the CdTe thin films deposited at substrate temperatures of 490°C and 520°C have transverse tensile stress, but at 550°C and 580°C have compressive stress.

The change of stress in the film may be caused by the combined effect of the different thermal expansion coefficients of the substrate and CdTe and the grain boundary being pulled by the grains during the cooling process. The substrate thermal expansion coefficient obtained from the supplier is 9.0×$\rm 10^{-6}$/K, while that of CdTe is 5.9×$\rm 10^{-6}$/K, which leads to the CdTe films deposited at higher substrate temperatures being more affected by the substrate contraction when cooled to room temperature, resulting in an increase of transverse compressive stress in the CdTe films. 

\begin{figure}[htp]
    \centering
    \includegraphics[width=12cm]{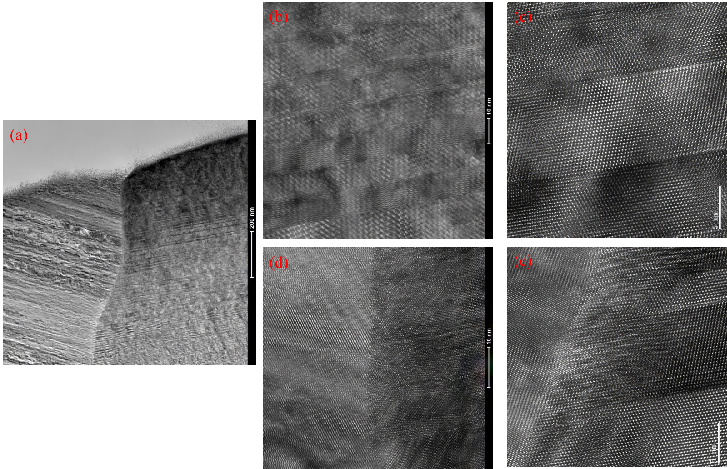}
    \caption{TEM image of the morphology near the grain boundary of the CdTe sample with a substrate temperature of 580°C.}
    \label{fig.5}
\end{figure}

\begin{figure}[htp]
    \centering
    \includegraphics[width=12cm]{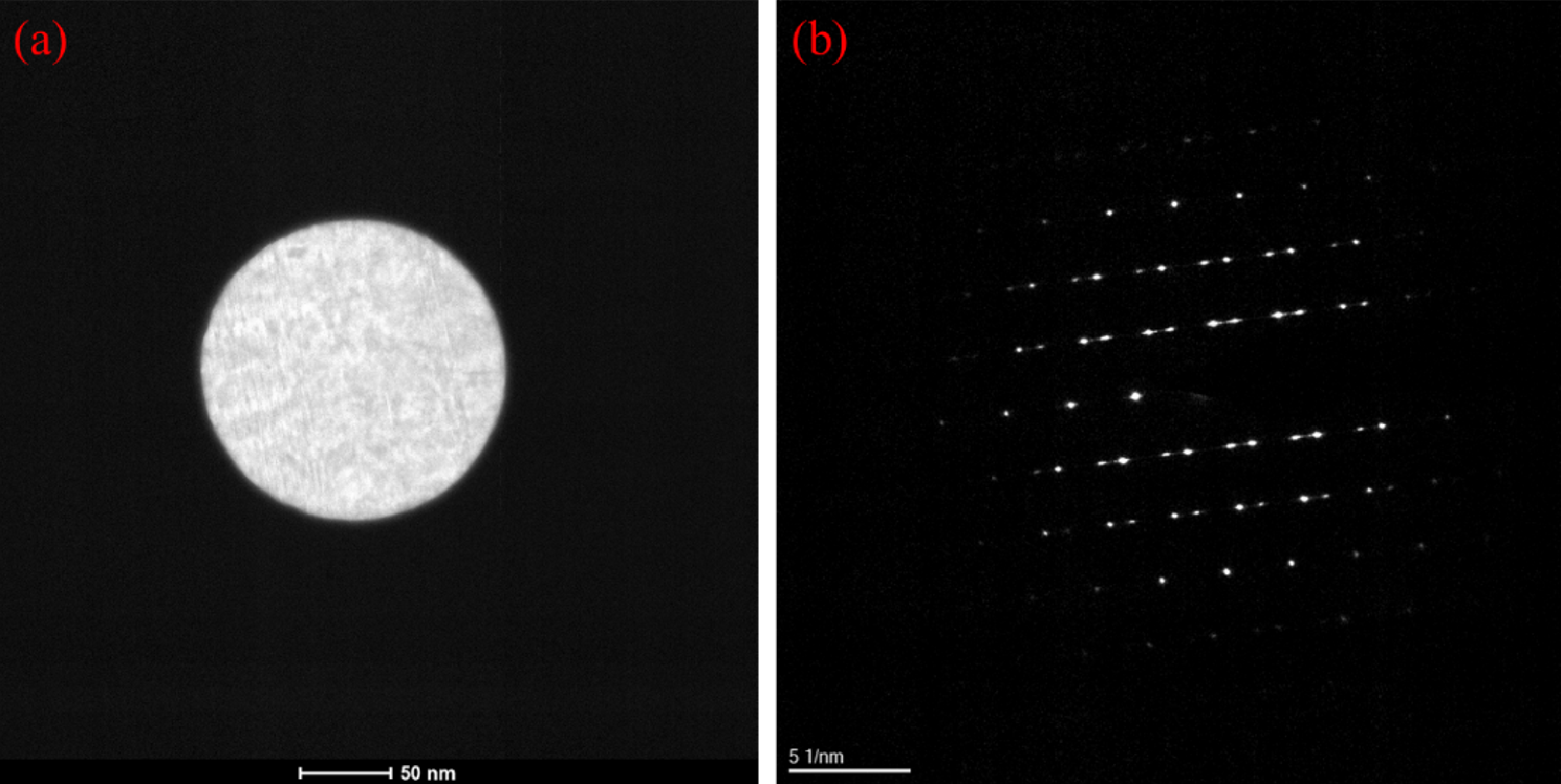}
    \caption{SAED selected area position and diffraction pattern of the twin position.}
    \label{fig.6}
\end{figure}

To further observe the internal situation of the film, FIB was used to cut vertically to the surface of the sample with a substrate temperature of 580°C, and the situation near the grain boundary was observed by TEM. Fig.\ref{fig.5}(b) and Fig.\ref{fig.5}(c) show the morphology of the twins inside the grain, and figure Fig.\ref{fig.5}(c) and Fig.\ref{fig.5}(d) show the morphology of the twins extending to the grain boundary. It can be observed that there are more stacking faults at the grain boundary. SAED was performed at the selected area of the observed twin position, as shown in Fig.\ref{fig.6}(a). From the diffraction pattern in Fig.\ref{fig.6}(b), paired diffraction spots can be seen, which are obvious characteristics of twins. In addition, there is a situation of diffraction spot dispersion and elongation, which indicates that there are stacking fault defects at both the grain boundary and the twin position, which ultimately affect the internal stress of the film.

The results indicate that the films deposited by GVD method at high temperature in this work increased the density and activity of dislocations in the crystal, promoted the formation and expansion of stacking faults, and the film began to form twins under the dual influence of thermal activation energy and stacking faults. The grown CdTe films belong to the cubic crystal system with a face-centered cubic lattice, which is easy to form intrinsic and extrinsic stacking faults. No matter what kind of stacking faults, the slip will gradually terminate at some point inside or on the surface of the crystal during the cooling process after the film growth, resulting in partial dislocations, which aggravate the lattice distortion at the grain boundaries.At the same time, it is observed from SEM that when the substrate temperature is 490°C and 520°C, the grain size is small and the grain boundaries are more, which leads to more atoms at the grain boundaries. In the process of cooling after depositing the film, the interplanar spacing in the crystal gradually recovers to room temperature state, and more atoms at the grain boundaries are subjected to tensile stress when the grains shrink, hindering the recovery of interplanar spacing. Combined with the calculation results of film stress, the films deposited at 490°C and 520°C have a greater influence of tensile stress at the grain boundaries. As the substrate temperature further increases, the difference of thermal expansion coefficient between substrate and CdTe dominates the influence of film stress.

In addition, we found that all twin interfaces point to grain boundaries, which is unfavorable for the preparation of battery devices, it is reported that the twin interface is a carrier transport channel\cite{ref11}, if the interface terminates at the grain boundary, it will cause carrier recombination, so Cd$\rm Cl_2$ annealing recrystallization becomes particularly important, if this situation cannot be changed, it will seriously affect the performance of CdTe solar cell devices.

\section{Summary}

In this work, CdTe films were grown on substrates with different temperatures using a self-made GVD device. As the substrate temperature increased, the grain size of the films showed an increasing trend. From the perspective of physical chemistry, the main reason was that the increase of substrate temperature was accompanied by the decrease of deposition rate, the solid-phase nucleation radius gradually increased, and finally led to the increase of grain size. The film orientation and stress were analyzed by XRD. As the substrate temperature increased, the (111) crystal plane orientation of the film weakened, and the other crystal plane orientations were enhanced to different degrees. The film showed a random orientation trend, which might be related to the lower nucleation barrier of the (111) crystal plane compared with other crystal planes. When the substrate temperature was high enough to reach the energy for other orientations to nucleate, the preferred orientation of the film became worse. However, for CdTe solar cells, the randomly oriented absorber layer was beneficial to light scattering and trapping, which was helpful for preparing high-quality solar cells. As the substrate temperature increased, the transverse stress of the film gradually changed from tensile stress to compressive stress. It was speculated that this might be the result of the combined effect of the thermal expansion coefficient difference between substrate and CdTe film and the stacking fault defects. Therefore, combining these two situations, seeking a suitable thermal expansion coefficient material as CdTe substrate would have a positive effect on reducing the stress influence during cooling process after CdTe film deposition at high temperature.

\end{document}